\documentclass[%
preprint,
 amsmath,amssymb,
 aps,
]{revtex4-2}

\usepackage{graphicx}
\usepackage{dcolumn}
\usepackage{bm}
\usepackage{tabularx}
\usepackage{epstopdf}
\usepackage{mathrsfs}

\usepackage{color}

\begin{document}


\title{Modeling Spatial Synchronization of Predator-Prey Oscillations via the XY Model under Demographic Stochasticity and Migration}

\author{Solmaz Golmohammadi}
\affiliation{%
Department of Physics, Institute for Advanced Studies in Basic Sciences (IASBS), Zanjan, 45137-66731, Iran
}%
\affiliation{
The Abdus Salam International Centre for Theoretical Physics (ICTP), Strada Costiera 11, 34014 Trieste, Italy
}%

\author{Mina Zarei}%
 \email{mina.zarei@iasbs.ac.ir}
\affiliation{%
Department of Physics, Institute for Advanced Studies in Basic Sciences (IASBS), Zanjan, 45137-66731, Iran
}%

\author{Jacopo Grilli}
 \email{jgrilli@ictp.it}
\affiliation{
Quantitative Life Sciences section, The Abdus Salam International Centre for Theoretical Physics (ICTP), Strada Costiera 11, 34014 Trieste, Italy
}%


\begin{abstract}
We investigate stochastic predator-prey dynamics and their spatial phase synchronization using the Rosenzweig-MacArthur model coupled across multiple patches. Combining stochastic simulations based on the Gillespie algorithm with analytical methods inspired by the XY model, we uncover fundamental mechanisms through which demographic noise and dispersal shape synchronization and phase transitions. This study offers a theoretical foundation for understanding and managing large-scale ecological synchrony and ecosystem resilience. \end{abstract}

\maketitle


\section{\label{sec:IIII}Introduction}

Population oscillations are a typical and well-documented phenomenon in ecological systems. These regular ups and downs in the number of individuals arise from the interplay of biological processes such as predator-prey interactions, resource variability, and environmental fluctuations. These cyclical population size changes have been observed in a wide variety of species and ecosystems, where they play an important role in shaping ecosystem dynamics~\cite{gilpin1973hares, Elton1942, butler1953nature, korpimaki2005predator, higgins1997stochastic, kamata2000population}.

A notable pattern that often emerges in this context is synchronization, in which populations in spatially separated habitats exhibit simultaneous fluctuations. Spatial synchrony, documented in both controlled laboratory settings~\cite{vasseur2009phase,fox2011phase} and in natural environments~\cite{Elton1942, Keith1963, Hanski1993, Sinclair1993, Royama1992, Moran1953, Bulmer1974, Korpimaki1996, Ranta1997c}, can have a major impact on ecosystem resilience~\cite{Liebhold2004}. When population cycles become tightly synchronized across regions (a state often referred as phase locking), the peaks and troughs of abundance tend to align in time. This coherence can make ecosystems more vulnerable to large-scale disturbances, such as extreme weather events, disease outbreaks, or resource shortages, potentially triggering widespread or system-wide collapses~\cite{Palmqvist1998,Abbott2011,MatterRoland2010,Schindler2010}.

A number of mechanisms have been proposed to explain how such synchrony arises. The Moran effect, for instance, links synchrony to correlated environmental variability across space. Another key process is dispersal (the movement of individuals between locations), which can physically couple populations and align their cycles~\cite{Moran1949, Moran1953, Zweigler1977, SinclairGosline1997, HaydonSteen1997, Blasius1999, Bjornstad1999a, Bjornstad1999b, Ranta1997a, Ranta1997b, Ranta1997c, Ranta1997d,vasseur2009phase}. Both these mechanisms explain why the synchrony typically decreases as geographic distance increases: since environmental correlations weaken, and long-distance dispersal becomes less common~\cite{Koenig2002, Lande1999, Ranta1995, Ranta1999, Sutcliffe1996, Paradis1999}.

Nevertheless, contrary to these expectations, synchrony has also been observed over remarkably large spatial scales~\cite{vasseur2009phase,fox2011phase}. Studies suggest that even limited, short-distance dispersal can be enough to synchronize predator-prey systems that inherently demonstrate cyclic fluctuations. In fact, minimal movements can phase-lock population cycles across an entire system, maintaining coordinated oscillations~\cite{fox2011phase, Blasius1999, Jansen1999, Jansen2001}. A classic example is the extensively studied synchronized population dynamics of snowshoe hares and their predators, particularly lynx, observed across multiple geographic scales~\cite{krebs2013synchrony, Blasius1999}. These findings are further supported by a variety of experimental and field studies~\cite{fox2011phase, comins1992spatial}.

A central debate in spatial ecology concerns the primary drivers of this synchrony. While correlated environmental fluctuations (the Moran effect) are a well-established mechanism ~\cite{Moran1949, Moran1953}, the role of dispersal as a synchronizing agent remains complex, especially given that local demographic noise tends to desynchronize patches. Early theoretical work highlighted that the efficacy of dispersal can be subtle. For instance, dispersal is a much more potent synchronizer when predator-prey dynamics exhibit a significant separation in their intrinsic time scales~\cite{goldwyn2008can}.

In recent years, frameworks from statistical physics have provided powerful new analogies. A number of studies have shown that the transition from incoherence to synchrony can be mapped onto the Ising model, especially in systems that show two-cycle oscillations. These works suggest that emergent, long-range synchrony can arise purely from local interactions, with the transition belonging to the Ising universality class~\cite{noble2015emergent,nareddy2020dynamical}.

Although environmental factors and species dispersal have been extensively studied in relation to spatial synchrony~\cite{fox2011phase, haydon2000spatial, haydon2001phase}, the role of demographic stochasticity is still not as well understood. This form of randomness comes from the random nature of individual births, deaths, and interactions, which happen as discrete events, and is particularly significant in predator-prey systems~\cite{black2012stochastic}. Previous research has demonstrated that demographic stochasticity plays a crucial role in population dynamics, influencing the stability of population oscillations, their frequency, and phase~\cite{golmohammadi2023effect}, and potentially causing extinction~\cite{smith2016extinction}. 

Despite extensive research on spatial synchrony driven by environmental variability and species dispersal, a critical knowledge gap remains: how does movement influence the phase relationships of limit cycles in spatially extended, stochastically oscillating populations? Most previous studies have focused either on purely deterministic models or on simulations that incorporate demographic stochasticity without rigorous analytical frameworks.

Our research fills this gap by combining discrete stochastic simulations based on the Gillespie algorithm~\cite{gillespie1977exact} with analytical approaches inspired by statistical physics. Drawing on the well-established XY model (a canonical framework for coupled phase oscillators and collective synchronization~\cite{plischke1994equilibrium, kardar2007statistical}), we qualitatively describe how demographic noise and movement interact to synchronize population cycles across spatial patches. By framing stochastic Rosenzweig-MacArthur predator-prey dynamics within the XY universality class, we uncover fundamental mechanisms of spatial phase synchronization. These insights not only advance our theoretical understanding of large-scale ecological synchrony but also can enhance our ability to predict and manage synchronized population collapses, ultimately supporting ecosystem resilience.

The paper is organized as follows: Section~\ref{sec:II} presents the methodology and analysis in three parts. First, we introduce the deterministic Rosenzweig-MacArthur (RMA) predator-prey model, identify parameter regimes exhibiting limit cycles, and describe how two patches are coupled via short-distance dispersal. Next, we consider the discrete version of the model to derive transition rates and apply them in stochastic simulations using the Gillespie algorithm. Finally, we analytically show that the coupled stochastic RMA model is mathematically equivalent to the XY model. We also confirm that the correlation lengths of the RMA model on one- and two-dimensional lattices match those known for the XY model. Section~\ref{sec:V} offers discussion and concluding remarks. 

\section{\label{sec:II}Methodology and Analysis }
\subsection{Deterministic RMA Model with Population-Gradient Migration}
The Rosenzweig-MacArthur Predator-Prey (RMA) model \cite{rosenzweig1963graphical} offers a standard framework for studying predator-prey dynamics. As a generalized extension of the Lotka-Volterra model \cite{lotka1920analytical} using a Holling type II response \cite{holling1959components}, it captures population oscillations observed in both natural and laboratory settings. The model describes the time evolution of prey (R) and predator (F) populations using the following equations:
\begin{eqnarray}
&& \dot{R} = aR-\dfrac{R^2}{2N}- \dfrac{sRF}{1+ s \tau R}, \nonumber \\  
&& \dot{F} = -d F+ \dfrac{sRF}{1+ s \tau R} \ .
\label{eq001}
\end{eqnarray} 

The population of preys exhibit logistic growth with rate $a$ and carrying capacity $2Na$ in the absence of predators ($F=0$), where $N$ indicates system size. In contrast, the population of predators declines with mortality rate $d$ if preys are not present ($R=0$). Predation follows a Holling type II response, which assumes a maximum per-capita predation rate at $1/\tau$ for large prey numbers, with $\tau$ being the handling time. The base attack rate is $s$.

By choosing the global timescale parameter $d=1$ and normalizing the population sizes as $r = \frac{R}{N}$ for preys and $f = \frac{F}{N}$ for predators, the model reduces to the following dimensionless form:
\begin{eqnarray}
&& \dot{r} = a r - \dfrac{r^{2}}{2} - \dfrac{\sigma r f}{1 + \sigma \tau r}, \nonumber \\
&& \dot{f} = -f + \dfrac{\sigma r f}{1 + \sigma \tau r} \ , 
\label{eqYYY}
\end{eqnarray}
where $\sigma = sN$. This classical model has been extensively studied \cite{rosenzweig1963graphical, doi:10.1137/0512047, smith2016extinction, smith_rosenzweig, cheng1981asymptotic, rosenzweig1963graphical}. We focus in the following parameter regime
\begin{equation}
0 < \tau < 1, \quad \sigma > \sigma_0 = \frac{1}{2 a (1 - \tau)} \ .
\label{eq:param_regime}
\end{equation}
Within this range, the system has three fixed points:
\begin{equation}
M_1 = (0, 0), \quad
M_2 = (2a, 0), \quad
M_3 = \left(\frac{1}{\sigma (1 - \tau)}, \frac{2 a \sigma (1 - \tau) - 1}{2 \sigma^2 (1 - \tau)^2}\right) \ ..
\label{eq:fixed_points}
\end{equation}
The equilibrium $M_3$ represents stable coexistence of prey and predator populations when
\begin{equation}
\sigma_0 < \sigma < \sigma^* = \frac{1 + \tau}{2 a \tau (1 - \tau)},
\label{eq:stability}
\end{equation}
beyond which $M_3$ becomes unstable and the system exhibits a stable limit cycle.
When a limit cycle emerges, it can be effectively approximated by an elliptical trajectory centered around the equilibrium point $M_3$, as described in \cite{golmohammadi2023effect}. This elliptical approximation gains accuracy as the ratio $\sigma/\sigma^*$ approaches unity from above.

The model described above assumes a well mixed population, where all the individuals interact with every other.
The effect of space can be incorporated in different forms. Here, we assume a discrete number of patches, within which the dynamics follow the well-mixed predator prey dynamics described above. Patches are coupled by migration~\cite{murdoch1992aggregation, elabdllaoui2002density, daoduc2009predator, holmes1994partial, molofsky1994population, mchich2005effects, kareiva1990population}.

In the simplest setting, we consider symmetric migration between two coupled predator-prey patches by adding explicit coupling terms to the population dynamics. For each patch \( j \in \{1, 2\} \), with \( k \) denoting the other patch (\( k \neq j \)), the prey and predator populations evolve according to
\begin{eqnarray} 
\dot{r}_{j} &=& a r_{j} - \frac{r_{j}^2}{2} - \frac{\sigma r_{j} f_{j}}{1 + \sigma \tau r_{i}} + \mu (r_{k} - r_{j}), \nonumber \\ 
\dot{f}_{j} &=& -f_{j} + \frac{\sigma r_{j} f_{j}}{1 + \sigma \tau r_{j}} + \mu (f_{k} - f_{j}) \ .
\label{eq003}
\end{eqnarray}
The migration terms \( \mu (r_k - r_j) \) and \( \mu (f_k - f_j) \) represent the symmetric exchange of prey and predator-individuals between patches at rate \( \mu \). This formulation captures both the local predator-prey interactions within each patch and the bidirectional migration between patches. Such coupling can significantly affect the synchronization and overall dynamics of populations distributed across spatially separated regions.
\subsection{Stochastic RMA Model with Population-Gradient Migration}Demographic stochasticity, arising from the inherently probabilistic nature of individual births and deaths, induces random fluctuations in population size that affect both the amplitude and frequency of oscillations. This effect is especially pronounced in finite populations, where chance events can significantly influence the system’s dynamics.
\subsubsection{Gillespie Simulation}
Consider two coupled habitat patches in the Rosenzweig–MacArthur (RMA) model. To incorporate demographic stochasticity, we simulate the system using discrete, individual-based stochastic simulations driven by the Gillespie algorithm~\cite{gillespie1977exact}. This method relies on explicitly defined transition rates describing probabilistic changes between different system states.

For each patch \( i \in \{1, 2\} \), with the other patch denoted by \( j \neq i \), the intrapatch interaction transition rates are given by
\begin{eqnarray}
 && T((F_j,R_j+1),(F_k,R_k)|(R_j,F_j), (R_k,F_k)) = a R_j, \nonumber \\
&& T((F_j+1,R_j-1),(F_k,R_k)|(R_j,F_j), (R_k,F_k)) = \dfrac{sR_jF_j}{1+ s \tau R_j} ,\nonumber \\
&& T((F_j,R_j-1),(F_k,R_k)|(R_j,F_j), (R_k,F_k)) = \dfrac{R_j^2}{2N} ,\nonumber \\
&& T((F_j-1,R_j),(F_k,R_k)|(R_j,F_j), (R_k,F_k)) = F_j \ ,
\label{rate1}
\end{eqnarray}
where \( T(C_f \mid C_i) \) denotes the transition rate from an initial configuration \( C_i \) to a final configuration \( C_f \). These rates correspond to processes such as prey reproduction, predation, prey competition, and predator mortality within the first patch. A analogous expressions apply for the second patch.

Migrations are transitions that alter population counts across patches. The migration of predators or preys from
patch \( k \) to patch \( j \) are respectively quantified by
\begin{align}
T((F_j,R_j+1),(F_k,R_k-1)|(R_j,F_j),(R_k,F_k)) &= \dfrac{\mu}{N} R_k, \nonumber\\
T((F_j+1,R_j),(F_k-1,R_k)|(R_j,F_j),(R_k,F_k)) &= \dfrac{\mu}{N} F_k \ .
\label{rate2}
\end{align}

\begin{figure}[h!]
\centering
\begin{minipage}{.45\hsize}
\includegraphics[width=\linewidth]{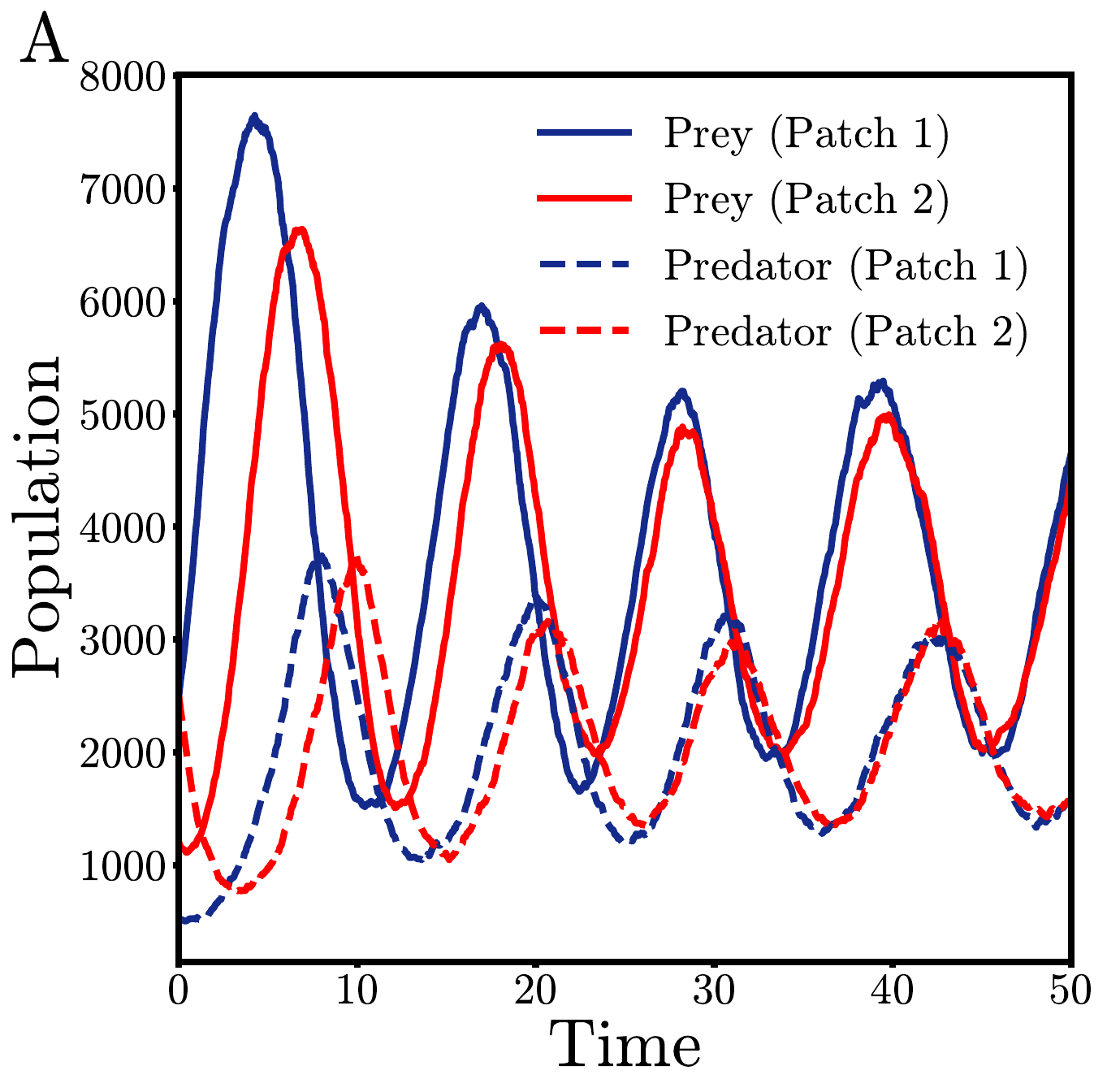}
\end{minipage}
\begin{minipage}{.45\hsize}
\includegraphics[width=\linewidth]{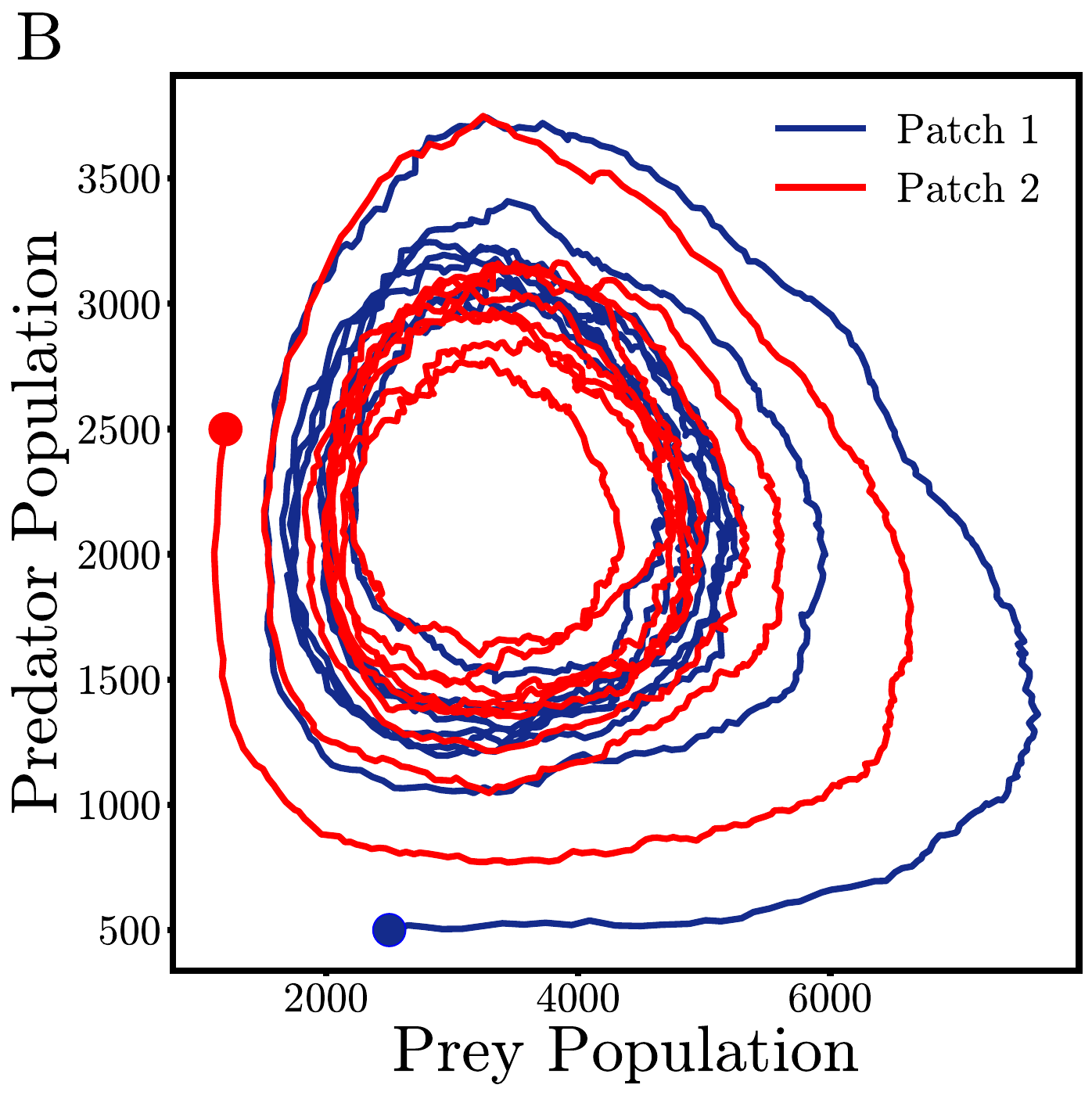}
\end{minipage}
\caption{Dynamics of two coupled stochastic patches in the Rosenzweig–MacArthur (RMA) model, distinguished by red and blue colors.
(A) Population trajectories over time: solid lines indicate prey dynamics, while dashed lines represent predator dynamics.
(B) Corresponding limit cycles in phase space, with bold points marking the initial states of the two patches.
Parameters used: $\tau = 0.5$, $a = 1$, $\gamma = 0.03$, $\mu = 0.01$, and $N = 5000$.
}
\label{fig:1}
\end{figure}

Using these rates, we perform stochastic simulations of the coupled RMA patches with the Gillespie algorithm. This algorithm explicitly simulates events --- births, deaths, and migrations --- one at a time, advancing time accordingly. The Gillespie algorithm efficiently captures demographic noise by avoiding unnecessary computations during inactive periods, thereby providing an accurate individual-based representation of predator-prey dynamics.

While the stability conditions in Eq.~\ref{eq:stability} address local stability in the deterministic model, demographic stochasticity significantly impacts long-term predator-prey dynamics. Random variations in birth, death, and migration create fluctuations around the equilibrium, making it metastable and ultimately leading to extinction in finite populations~\cite{smith2016extinction}. Here, we focus on large populations and the transient behavior along the limit cycle before extinction occurs.

Figure~\ref{fig:1} illustrates the resulting stochastic population oscillations and the limit cycle behavior of two coupled RMA systems. Although each patch starts with different initial conditions, their phases tend to align or synchronize over time as a result of the interaction term. Given that the system exhibits a stable limit cycle at steady state, we can focus on the phase variables of each patch rather than tracking the prey and predator populations directly, allowing us to construct a stochastic model of the system using Langevin dynamics.
\subsubsection{\label{analytical}Langevin and Angular Brownian Modeling}

\begin{figure}[h!]
\centering
\begin{minipage}{.45\hsize}
\includegraphics[width=\linewidth]{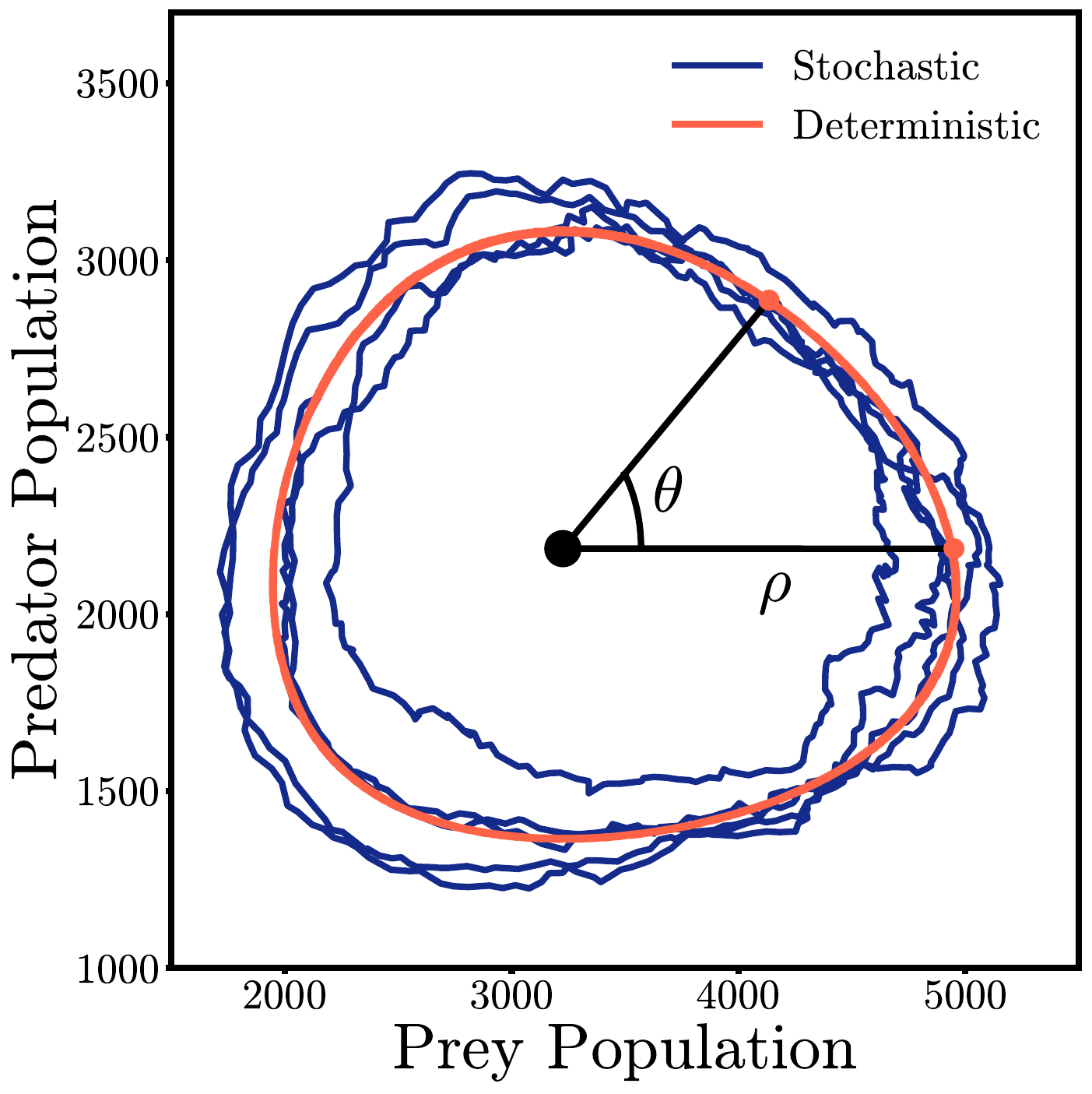}
\end{minipage}

\caption{Limit cycles of the deterministic and stochastic RMA models for a single, uncoupled patch. The phase space is represented in terms of radius ($\rho$) and phase ($\theta$), with the deterministic limit cycle shown in red. Parameters used are $\tau = 0.5$, $a = 1$, $N = 5000$, and $\sigma = 3.1$.}
\label{fig:2}
\end{figure}

 For a single uncoupled patch, the limit cycle trajectory is well approximated by an ellipse when its typical amplitude is small, i.e. close to the Hopf bifurcation~\cite{golmohammadi2023effect}. This enables the analytical characterization of the oscillating trajectory using a phase variable \(\theta(t)\) (illustrated in Figure~\ref{fig:2}). Fluctuations in the phase’s oscillation frequency can be effectively modeled as angular Brownian motion, described by the following Langevin equation:
\begin{eqnarray}
\dot{\theta} = \omega + \sqrt{D} \xi(t) \ .
\label{eq8}
\end{eqnarray}
Here, $\omega$ refers to the deterministic frequency, which is defined by the RMA model parameters $a$, $\tau$, and $\sigma$; $D$ indicates the noise strength, which can be estimated analytically or via stochastic simulations of the RMA model and is influenced by the system size along with other parameters~\cite{golmohammadi2023effect}. Additionally, $\xi(t)$ represents Gaussian white noise with zero mean and delta-correlated fluctuations.

To incorporate interactions between patches into this stochastic framework, the Langevin equation can be appropriately extended. In the case of two coupled stochastic RMA oscillators, the phase evolution of each patch is given by:
\begin{eqnarray}
\dot{\theta}_{i}(t) = \omega + \mu F_{i}(\theta_{i}, \theta_{j}) + \sqrt{D}\xi_{i}(t),
\label{eq004}
\end{eqnarray}
where the indices \( i, j \in \{1, 2\} \) with \( j \neq i \) label the two patches. In this equation, $\mu$ denotes the coupling strength, and the key task is to determine the interaction function $F_i(\theta_i, \theta_j)$, which plays a central role in defining the effective Hamiltonian for the coupled patch system.

To derive a phase-based description of the system, we first expand the population density Eq.~\ref{eq003} around the equilibrium point located at the center of the limit cycle. We then apply an elliptical transformation defined by $dr = \rho \cos(\theta)$ and $df = \rho \epsilon \sin(\theta)$, where $\rho$ and $\rho\epsilon$ represent the semi-major and semi-minor axes of the elliptical orbit. This transformation yields the following expression for the interaction term between patches (see Supplementary Material for derivation):
\begin{eqnarray}
F_{i} = -\mu \sin(\theta_{j} - \theta_{i}).
\label{eq006}
\end{eqnarray}
Defining the phase difference as $\phi = \theta_j - \theta_i$, and assuming identical intrinsic frequencies $\omega$ and noise strengths $D$ for both patches --- since these are set by the system parameters~\cite{golmohammadi2023effect} --- we obtain the following stochastic differential equation governing the phase difference:
\begin{align}
\dot{\phi}(t) = -2\mu \sin\phi + \sqrt{2D} \xi(t) \ .
\label{langevin}
\end{align}
Here, we use the fact that if $\xi_1(t)$ and $\xi_2(t)$ are independent Gaussian white noise processes with zero mean and noise strength $D$, then their difference $\xi(t) = \xi_1(t) - \xi_2(t)$ is also Gaussian with zero mean and an effective variance of $2D$.  
The degree of synchrony between the two patches is characterized by the phase correlation metric 
$x = \cos \phi$. Applying this variable transformation, Eq.~\ref{langevin} can be rewritten as:
\begin{eqnarray}
\dot{x} = 2\mu (1-x^{2})- Dx+ \sqrt{1-x^{2}}\sqrt{2D}\xi(t) .
\label{eq10}
\end{eqnarray}
To determine the stationary distribution associated with this Langevin equation, we convert the stochastic differential equation into its equivalent Fokker-Planck partial differential equation, which governs the temporal evolution of the probability density function. The stationary distribution is obtained as the steady-state solution of the Fokker-Planck equation, where the probability density becomes time-independent, reflecting a balance between deterministic drift and stochastic diffusion components. The resulting stationary probability density function for the pairwise phase correlation metric is given by the following expression (a detailed derivation is provided in the Supplementary Material):
\begin{equation}
P_{s}(x) = \frac{Z}{2 D \sqrt{1 - x^{2}}} \exp\left( 2 \frac{\mu}{D} x \right),
\label{eq:stationary_prob}
\end{equation}
where \( Z \) is the normalization constant (which can be expressed as function of \(\mu\), see supplementary material). 

The stationary distribution \( P_s(x) \) can be written in exponential form, resembling Boltzmann-like distribution, expressed as
\[
P_s(x) \propto e^{-\beta H(x)},
\]
where \( H(x) \) denotes an effective Hamiltonian and \( \beta \) represents the inverse temperature parameter. Accordingly, the effective Hamiltonian for the system is defined by
\begin{equation}
\beta H(x) = 2 \frac{\mu}{D} x,
\label{hamiltonian}
\end{equation}
with \( \beta = \frac{2\mu}{D} \). Here, the noise intensity \( D \) acts as a temperature-like parameter that governs the magnitude of stochastic fluctuations. The multiplicative pre-factor \( \frac{1}{\sqrt{1 - x^{2}}} \) results from the Jacobian of the variable transformation \( x = \cos \phi \) and reflects the geometric structure of the phase space.

\begin{figure}
\begin{minipage}{.45\hsize}
\includegraphics[width=\linewidth]{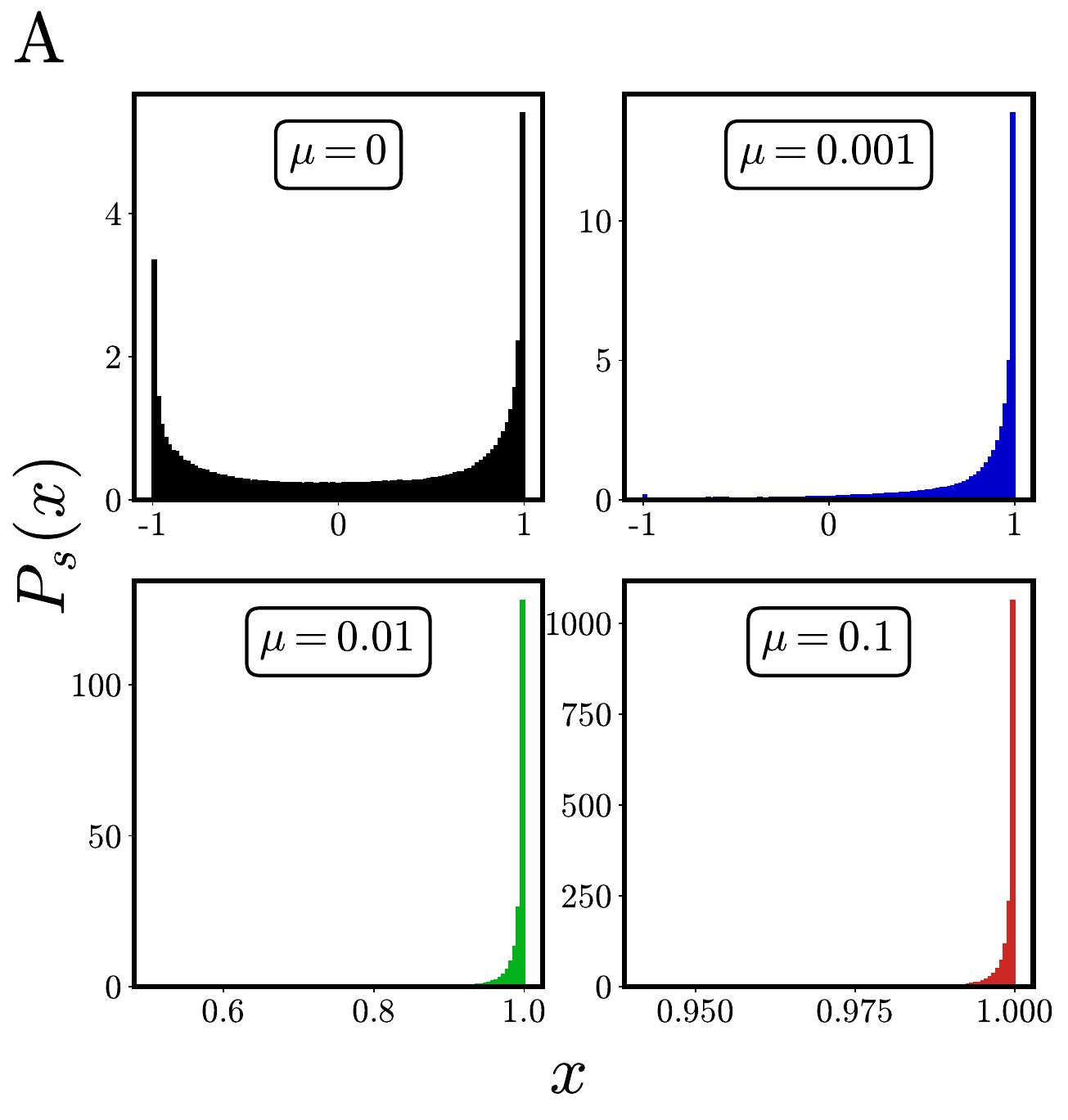}
\end{minipage}
\begin{minipage}{.47\hsize}
\includegraphics[width=\linewidth]{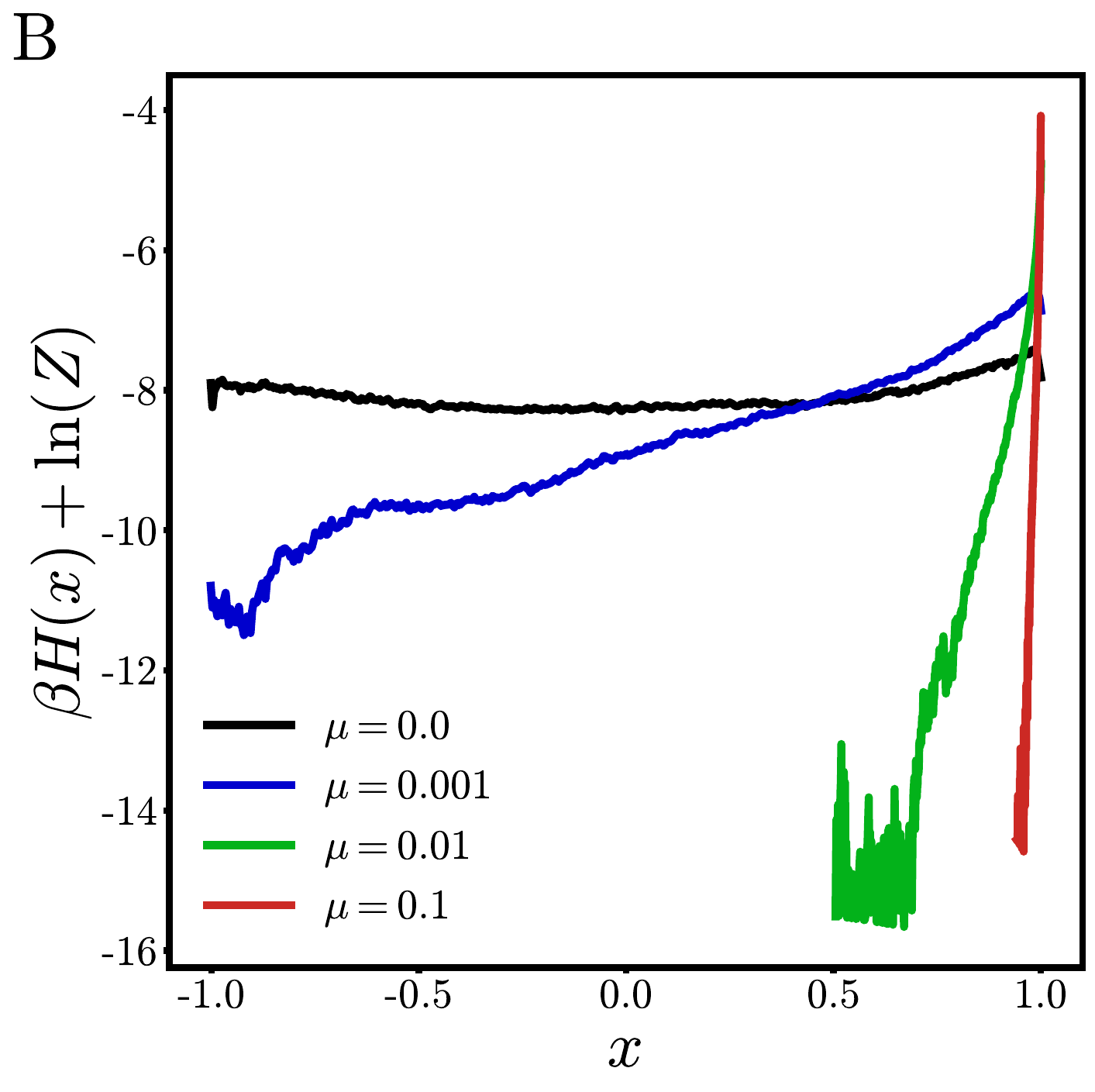}
\end{minipage}
\caption{Statistical characterization of phase synchrony in stationary population oscillations of two coupled patches.
(A) Histogram of the synchrony metric, defined as the cosine of phase differences in the stationary regime (denoted by $x$), shown for different migration constants.
(B) The Hamiltonian-equivalent term as a function of $x$, obtained by fitting the histogram data to the Boltzmann distribution.
Parameter values are $\tau = 0.5$, $a = 1$, $\gamma = 0.03$, and $N = 50000$. 
}
\label{fig:3}
\end{figure}

To validate the model and confirm the linear dependence of the Hamiltonian on \( x \), we first determine the stationary distribution \( p_s(x) \) via stochastic simulations of two coupled patches using the Gillespie algorithm. By substituting this distribution into Eq.~\ref{eq:stationary_prob}, we then extract the effective Hamiltonian --- calculating $\log p_s(x)$ --- and compare its slope to that predicted by the theoretical model. Figure~\ref{fig:3}(A) shows the stationary distribution of the phase correlation metric for different values of the coupling constant \( \mu \). It is evident that the distribution varies with the coupling strength. In the absence of coupling (\( \mu = 0 \)), the variable \( x = \cos \phi \) exhibits a characteristic U-shaped distribution over the interval \([-1, 1]\), which corresponds exactly to the situation where the phase difference \( \phi \) is uniformly distributed between \( -\pi \) and \( \pi \). This uniform distribution of phase differences in the stationary state implies that the patches' phases evolve independently. As the coupling constant increases, the distribution progressively becomes sharply concentrated around \( x = 1 \) (\( \phi = 0 \)), reflecting stronger phase synchronization between the patches.

By using the stationary distribution \( p_s(x) \) obtained from simulations together with Eq.~\ref{eq:stationary_prob}, we extract the effective Hamiltonian \( H(x) \) from the data. Figure~\ref{fig:3}(B) displays a plot of \( \ln Z - \beta H(x) \) based on these simulation results. The figure reveals a clear linear trend, with a slope that coincides with the theoretical prediction of \( \mu / D \). This agreement confirms that the effective Hamiltonian proposed by our model, \( \beta H(x) = (\mu / D)x \), is consistent with the simulation data.

\subsection{\label{subIII}Mapping Stochastic Coupled RMA Patches to the XY Model}

Our results show that the Hamiltonian for the two coupled patches (Eq.~\ref{hamiltonian}) depends linearly on \( x \), the cosine of their phase difference. This resembles the XY spin model~\cite{plischke1994equilibrium, kardar2007statistical} --- a classic model in statistical physics that describe planar spins whose Hamiltonian is given by
\begin{eqnarray}
\mathscr{H} = -J\sum_{\langle i, j \rangle} \cos (\theta_{i} - \theta_{j}),
\end{eqnarray}
where \(\theta\) denotes the spin angle relative to the x-axis, and the sum runs over neighboring pairs. This form captures an interaction energy that depends on the allignment of the spins. 

This result suggests that spatial phase synchronization observed in a structured network of patches may exhibit phenomena similar to those found in the XY model, including phase ordering and phase transitions.
The behavior of the XY model, including its phase transitions and the decay of spin correlations with distance, is strongly determined by the lattice’s dimensionality and topology. The spin-spin correlation function in the XY model is defined as  
\begin{eqnarray}
C(l) = \langle \cos(\theta_i - \theta_j) \rangle,
\end{eqnarray}
where \(\theta_i\) and \(\theta_j\) are the spin angles at lattice sites \(i\) and \(j\), separated by distance \(l = |i - j|\). In one-dimensional rings (periodic chains), strong thermal fluctuations prevent spontaneous symmetry breaking, so the classical XY model shows no long-range order or phase transition at finite temperatures. Consequently, Spin correlations decay rapidly with distance, as described by:
\begin{eqnarray}
C(l) = \langle \cos(\theta_i - \theta_{i+l}) \rangle \sim e^{-l/\xi},
\label{oneD}
\end{eqnarray}
where $(\xi \propto J/T)$ denotes the correlation length. Consequently, no long-range magnetic order emerges in the system.

In contrast, on two-dimensional periodic lattices (such as square lattices), the XY model undergoes a Berezinskii-Kosterlitz-Thouless (BKT) transition. Below the critical temperature vortex-antivortex pairs remain bound, resulting in quasi-long-range order characterized by spin correlations that decay algebraically as a power law:
\begin{eqnarray}
C(l) \sim l^{-\frac{T}{2 \pi J}}.
\label{twod1}
\end{eqnarray}
Above the critical temperature, these pairs unbind and proliferate, disrupting order and causing spin correlations to decay exponentially:
\begin{eqnarray}
C(l) \sim e^{-l/\xi}.
\label{twod2}
\end{eqnarray}
This exponential decay indicates a disordered phase.

We examine the mapping of the stochastic coupled RMA model onto the classical XY model by constructing two network configurations: a one-dimensional ring and a two-dimensional lattice, both with periodic boundary conditions. Stochastic simulations are carried out using the Gillespie algorithm, with interaction rates between patches calculated according to Eqs.~\ref{rate1} and \ref{rate2}. The spatial correlation lengths derived from these simulations are systematically compared with the well-established correlation length behavior characteristic of the XY model.

\begin{figure}[h!]
\begin{minipage}{.45\hsize}
\includegraphics[width=\linewidth]{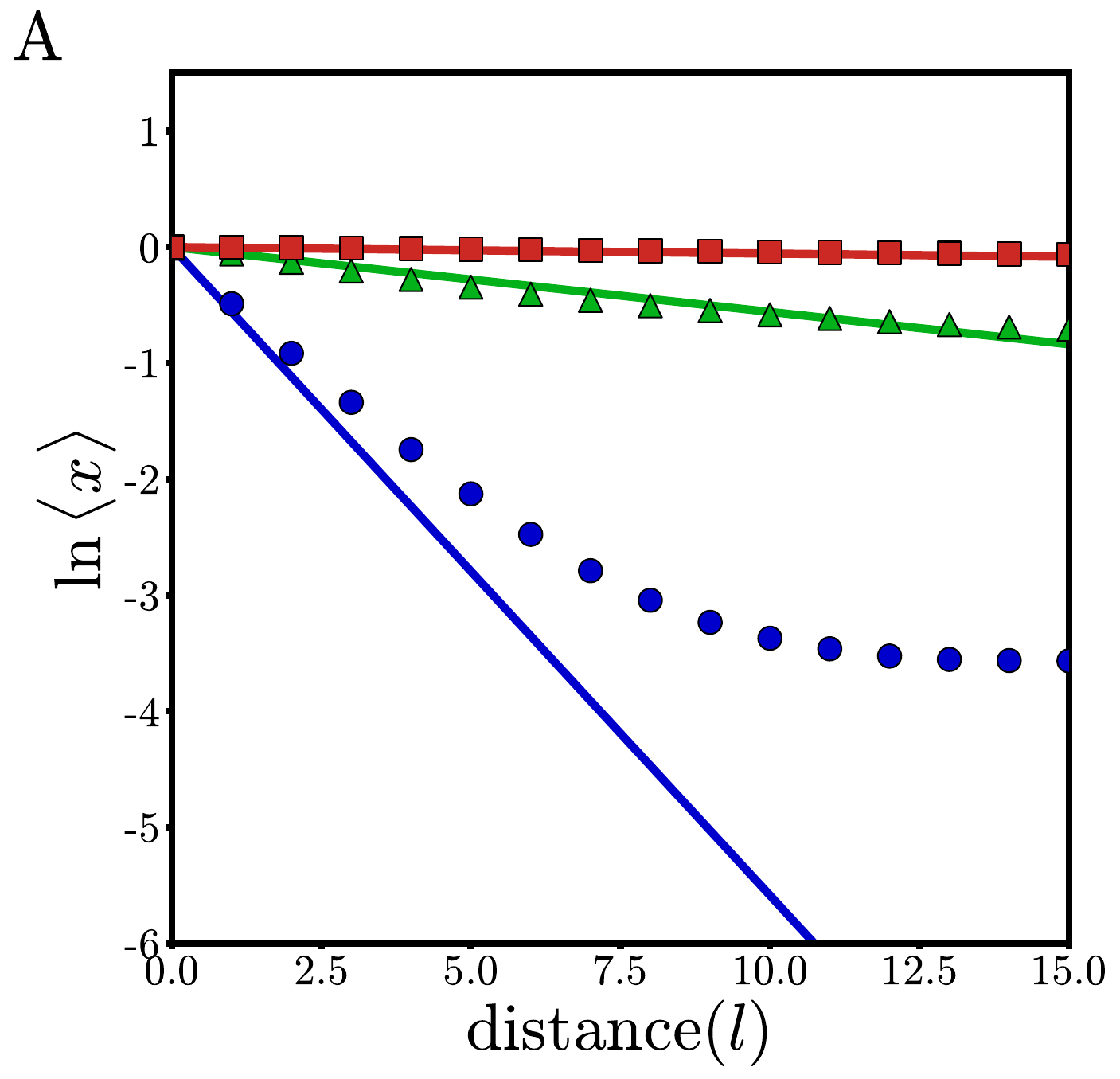}
\end{minipage}
\begin{minipage}{.45\hsize}
\includegraphics[width=\linewidth]{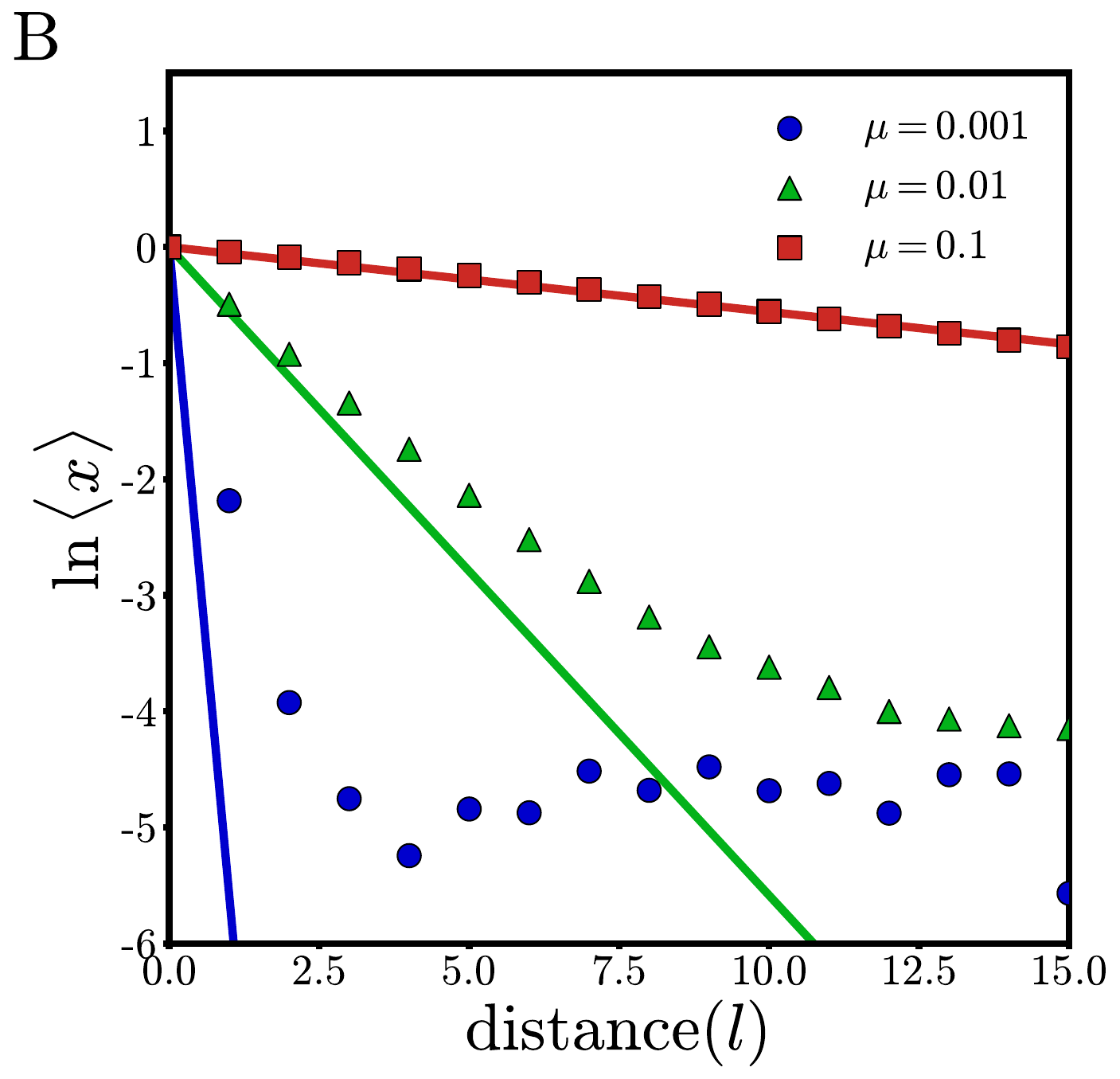}
\end{minipage}
\caption{Dependence of pairwise phase synchrony on inter-patch distance and noise in a ring network composed of 60 patches.
(A) Natural logarithm of the mean synchrony metric as a function of distance for (A) low noise ($N=50000$, $D = 0.00056$) and (B) high noise ($N=5000$, $D = 0.0056$). Averages are calculated over stationary intervals across 100 independent simulations and all pairs of patches with the same distance. Solid lines indicate the predictions from the XY model correlation function, $\ln C(l) = -\frac{D}{\mu}l$.
Parameters: $\tau = 0.5$, $a = 1$, $\gamma = 0.03$.}
\label{fig:4}
\end{figure}

\begin{figure}[h!]
\begin{minipage}{.45\hsize}
\includegraphics[width=\linewidth]{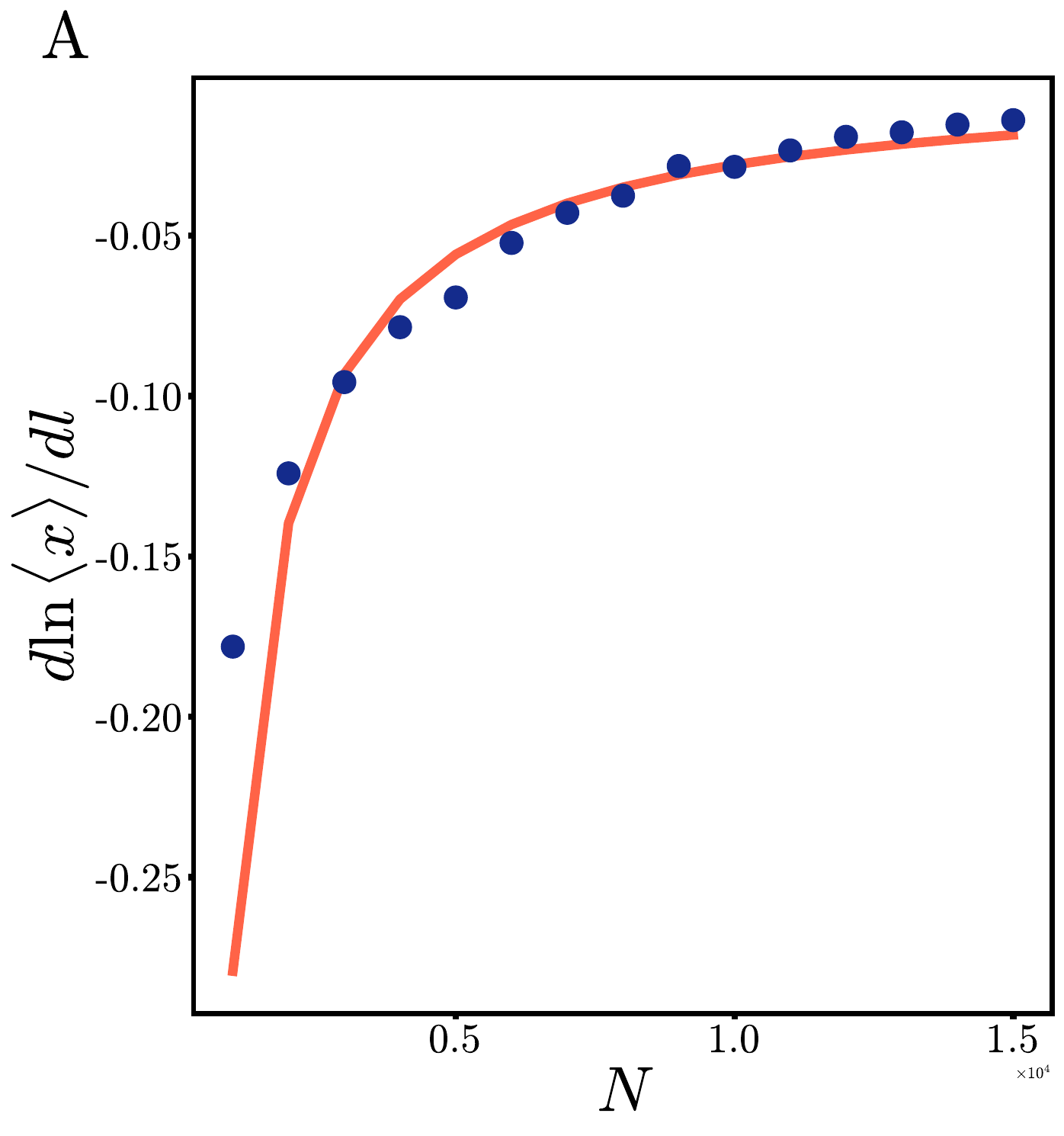}
\end{minipage}
\begin{minipage}{.45\hsize}
\includegraphics[width=\linewidth]{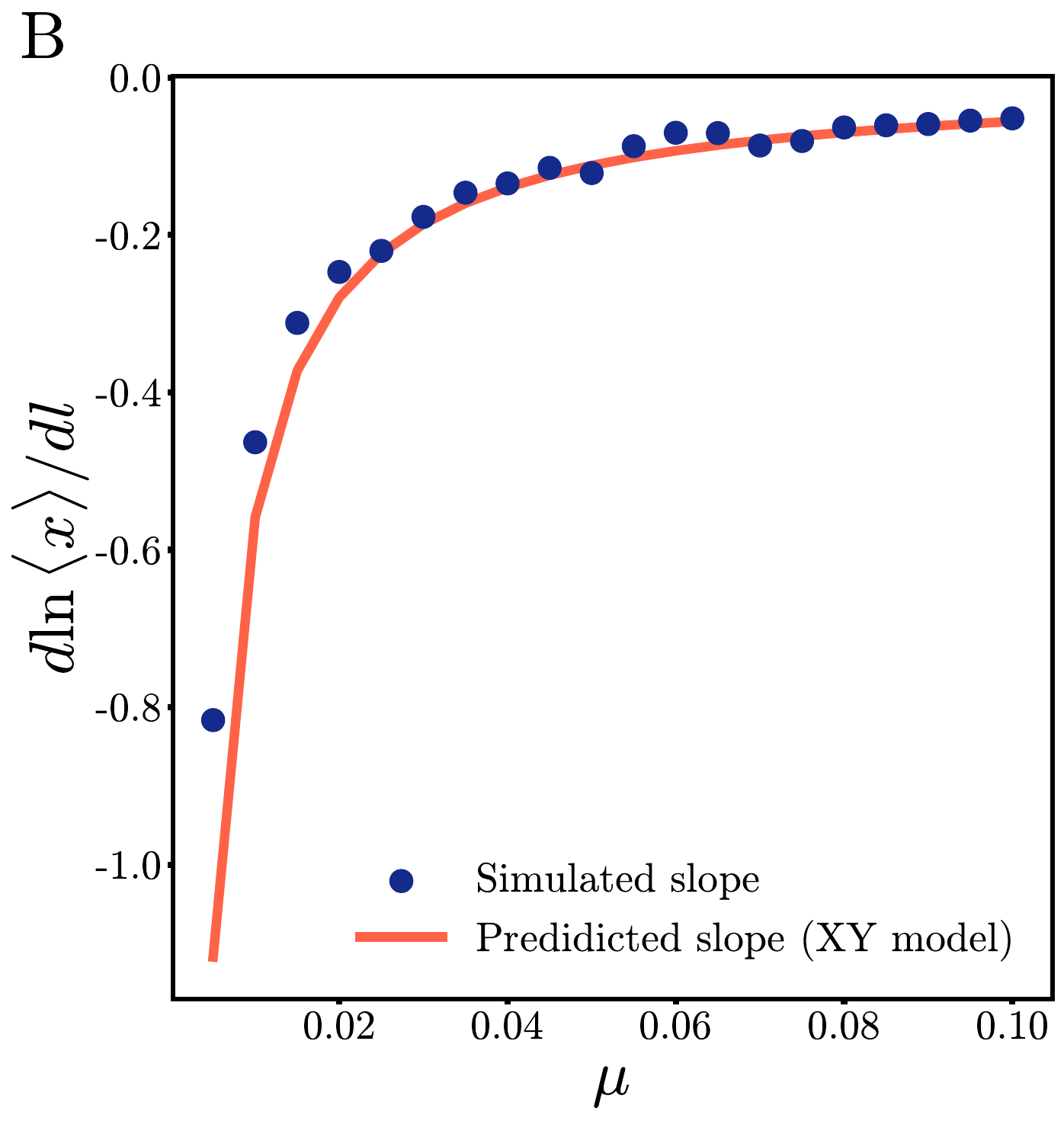}
\end{minipage}
\caption{Comparison of decay slopes in spatial correlations between RMA simulations and XY model predictions for 60 patches arranged in a ring, examined across (A) varying system sizes and (B) coupling strengths. Averages were computed over stationary time intervals from 100 independent simulation runs, considering all pairs of patches separated by the same distance. Solid lines represent the analytical prediction given by ($-Dl / \mu$). The parameters used in these figures are $N = 5000$, $D = 0.0056$, $\tau = 0.5$, $a = 1$, and $\gamma = 0.03$.}
\label{fig:5}
\end{figure}

Specifically, we analyze a one-dimensional ring consisting of 60 stochastic RMA patches, where each species migrates left or right with equal probability, to assess the accuracy of the XY model predictions. Figure~\ref{fig:4} shows the phase correlation, $C(l) = \langle \cos \phi \rangle$, as a function of distance $l$ for various coupling constants $\mu$. These results are averaged over stationary time intervals from 100 independent simulation runs, considering all patch pairs separated by the same distance. The key distinction between the two panels is the system size, which inversely modulates noise intensity. We observe that the decay of correlation closely follows the XY model’s predicted scaling determined by the ratio $\mu/D$ (see Eq.~\ref{oneD}). For both panels, the data exhibit excellent agreement with the predicted behavior over long distances when the $\mu/D$ ratio is large, corresponding to extended correlation lengths. This demonstrates that the correlation length is primarily controlled by the interplay between deterministic phase dynamics ($\mu$) and noise ($D$).
To extend our analysis over a broader range of parameters, we calculate and plot the slope of the natural logarithm of the correlation function with distance for different system sizes $N$ and coupling strengths $\mu$, as shown in Fig.~\ref{fig:5}. In this figure, the dots indicate slope values obtained from simulations, while the solid line represents the theoretical prediction $D/\mu$. The results show that the theoretical model provides a robust approximation over a wide range of parameters, particularly for large system sizes (corresponding to low noise levels) and stronger coupling. Under these conditions—when the correlation length is long—the simulation data agree closely with the analytical prediction.

In the two-dimensional setup, predator-prey patches are arranged on a square lattice with periodic boundary conditions, forming a toroidal topology. Within this stochastic RMA model framework, each species migrates randomly in four directions at rates defined by Eqs.~\ref{rate1} and \ref{rate2}. We calculate the average correlation length between pairs of patches as a function of their separation distance, where distances are measured using the Euclidean metric on the torus lattice. This averaging is performed over stationary time intervals and aggregated from 100 independent simulation runs, considering all patch pairs at equivalent distances. 

\begin{figure}[h!]
\begin{minipage}{.30\hsize}
\includegraphics[width=\linewidth]{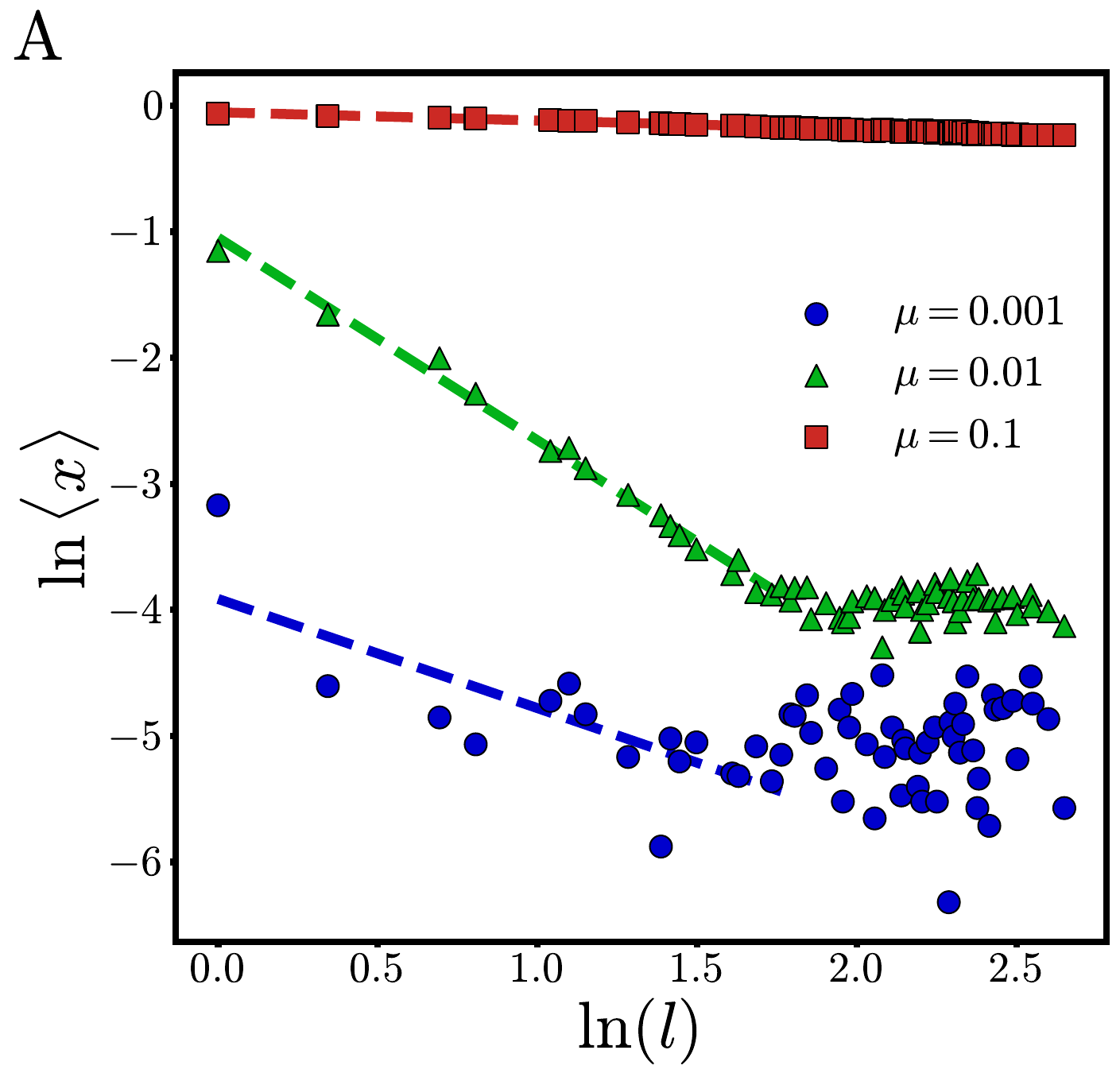}
\end{minipage}
\begin{minipage}{.30\hsize}
\includegraphics[width=\linewidth]{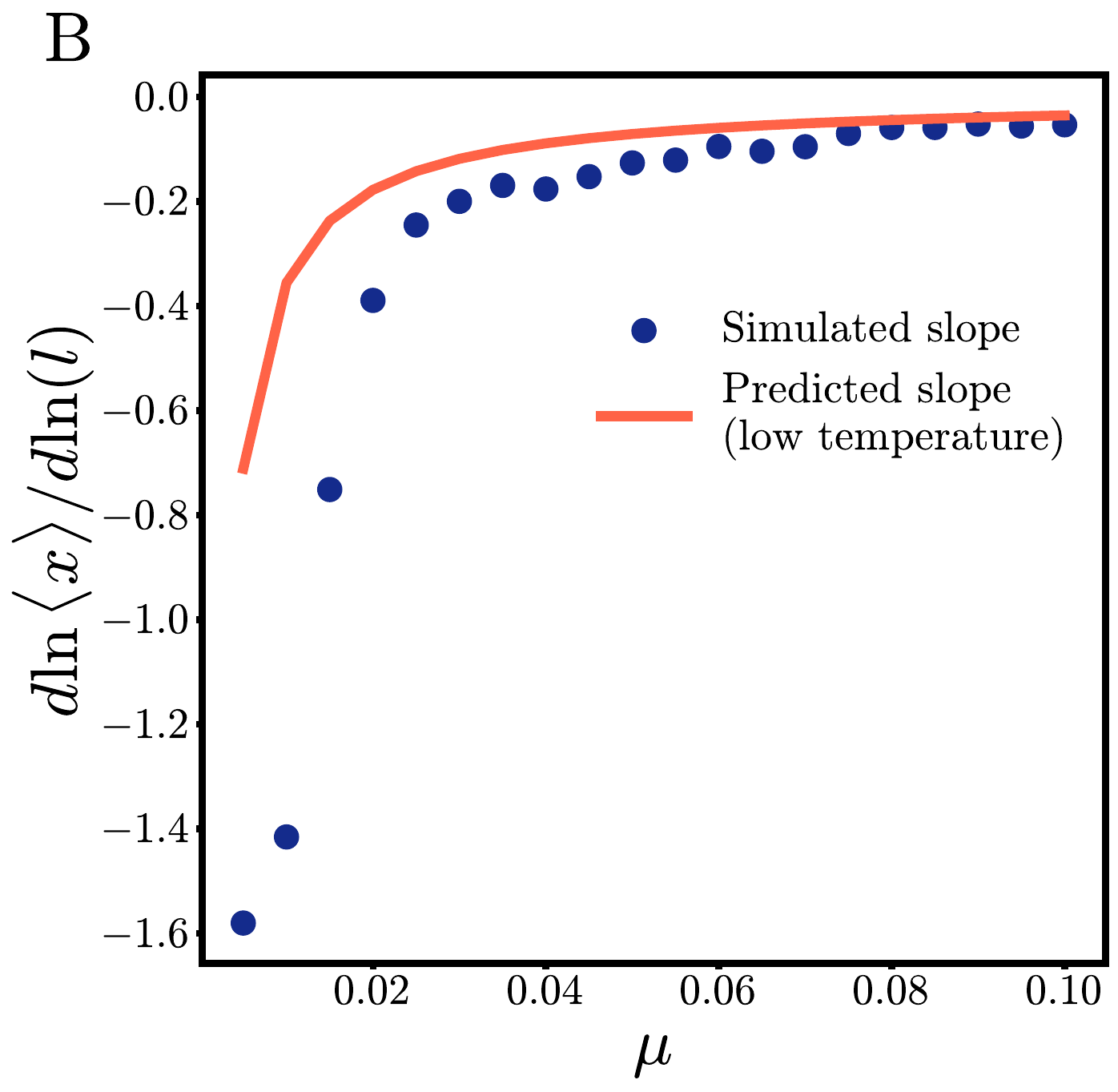}
\end{minipage}
\begin{minipage}{.31\hsize}
\includegraphics[width=\linewidth]{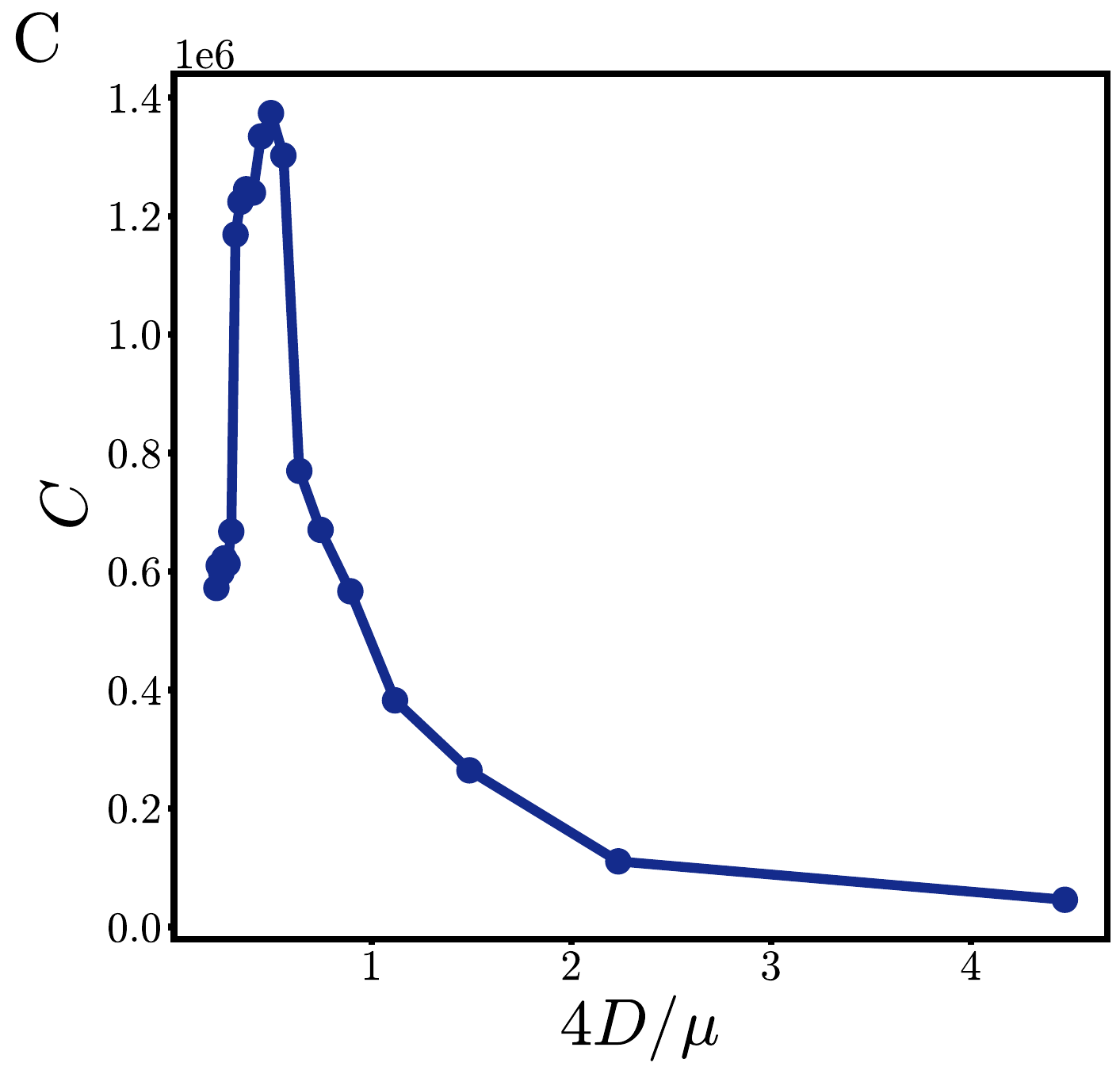}
\end{minipage}
\caption{Comparison of decay slopes in spatial correlations from stochastic RMA simulations and XY model predictions on a $20 \times 20$ periodic lattice.
(A) Pairwise synchrony plotted against patch distance, with the dashed line indicating the fit to the decaying region of the data.
(B) Fitted slope values across different coupling strengths. The solid lines indicate the analytical slope prediction of the XY model at low temperature, expressed as ($2D/\pi \mu$). (C) Specific heat, calculated from the energy variance, plotted as a function of $4D/\mu$. Data are averaged over stationary intervals from 100 independent simulations, including all patch pairs at equal distances.
Parameters: $N = 5000$, $D = 0.0056$, $\tau = 0.5$, $a = 1$, and $\gamma = 0.03$.}
\label{fig:6}
\end{figure}

Fig.~\ref{fig:6}A shows the correlation as function of distance for various coupling constants $\mu$. In these plots, points correspond to simulation results, while dashed lines represent linear fits applied to the decaying portion of the data before saturation.
Figure~\ref{fig:6} reveals two distinct regimes. At high values of the ratio $\mu/D$, the correlation function exhibits a power-law decay reminiscent of the two-dimensional XY model below its critical temperature. Importantly, the exponent of the power-law behavior decreases as $\mu/D$ increases, indicating a non-universal behavior, with slower decay and thus longer correlation. This behavior aligns with XY model predictions (see Eq.~\ref{twod1}). Improvements in fit quality are also observed when lowering $D$ or increasing system size, consistent with the interpretation of $D$ as an effective temperature parameter. 
Conversely, at low $\mu$ values,the correlation function does not follow a power-law decay. Instead, it decays exponentially, resembling the behavior of the two-dimensional XY model above its critical temperature.

Moreover, we compared the slopes of the natural logarithm of the pairwise correlation function from stochastic simulations with the analytical prediction of the XY model, $(2D/\pi \mu )$, over a wide range of $(\mu/D)$ values. Figure~\ref{fig:6}B shows that simulation results at higher $(\mu/D)$ closely match the XY model’s predictions for lower effective temperatures. However, at low $(\mu/D)$, deviations from the power-law scaling occur due to the  Berezinskii–Kosterlitz–Thouless (BKT) phase transition observed in the XY model, corresponding to the system crossing above the critical temperature. For high-migration rates and high-population sizes (low-noise), the system display long-range correlations resembling spatial synchrony, as observed in the low-temperature XY model. Decreasing the migration rate or increasing the strength of demographic noise makes the system cross a phase transition point and become spatially asynchronous, displaying a finite correlation length, as observed in the high-temperature phase of the XY model.

The specific heat, obtained from the effective energy variance, is plotted as a function of $4D/\mu$ in Fig.~\ref{fig:6}C. As expected for the XY model, a distinct peak emerges, clearly indicating the phase transition. The slight discrepancy observed in the critical temperature is likely due to finite-size effects, which cause the peak to shift away from the true critical temperature value in the thermodynamic limit. According to finite-size scaling theory, as the system size increases, the peak becomes sharper, and the measured critical temperature converges to the bulk value.
\section{\label{sec:V} Discussion}

Our main goal was to explore how short-range dispersal affects phase synchronization among predator-prey oscillators described by the Rosenzweig-MacArthur (RMA) system, drawing insights from the XY spin model—an established paradigm in statistical physics for understanding collective behavior.

A key methodological innovation was to approximate the population limit cycle as an elliptical trajectory centered on the co-existence fixed point~\cite{golmohammadi2023effect}. This simplification made it possible to reduce the system's two-dimensional oscillations to a single phase variable, emphasizing the phase differences that drive synchronization patterns across spatial patches. By assuming a constant oscillation radius (an approach supported by prior studies), we decoupled amplitude fluctuations from phase dynamics, which further simplified the analysis.

We incorporated demographic stochasticity (parameterized by system size), which added noise into the phase equations. The interplay between stochastic fluctuations and deterministic coupling revealed patterns analogous to the BKT transition in the two-dimensional XY model. Larger populations, with reduced demographic noise, tended to exhibit phase synchronization dominated by dispersal and coupling. In contrast, smaller and noisier systems tended to lose synchronization. This balance highlights the critical thresholds where large-scale phase coherence emerges or breaks, affecting the resilience of the ecosystem.

It is useful to compare our XY model-based approach with the recent successful application of the Ising model to ecological synchrony~\cite{noble2015emergent,nareddy2020dynamical}. These studies powerfully demonstrated that the transition to synchrony in populations with discrete, two-cycle dynamics falls within the Ising universality class. This was a significant finding, linking emergent ecological order to a fundamental principle of critical phenomena. However, the Ising model, which describes discrete spin states (e.g., ``up'' or ``down''), is less naturally suited to the continuous-phase oscillations characteristic of many predator-prey systems, such as the limit cycles produced by the Rosenzweig-MacArthur model. The XY model, by describing rotors with a continuous phase angle from $0$ to $2\pi$, provides a more analogous framework for these systems. Our findings, therefore, complement the Ising-based work, suggesting that the choice of statistical mechanics model should be dictated by the underlying nature of the ecological oscillator itself: discrete two-state cycles may map to the Ising model, while continuous limit cycles map more naturally to the XY model.

High synchronization reduces population variability across landscapes(their fluctuations across different landscapes start to look very similar). As a result, the overall resilience of the ecosystem to environmental perturbations tends to decline. Understanding the mechanisms driving such synchronization is therefore essential for predicting ecological responses and designing effective conservation strategies.

In this context, framing the stochastic RMA model within the same universality class as the XY model is a significant advance. It makes it possible to use well-established theoretical tools from statistical physics (such as phase transition theory and critical exponents) to study ecological systems. Beyond empirical or simulation-based work, this link provides a more general framework for exploring how local stochasticity and dispersal shape large-scale patterns. Ultimately, integrating ecological dynamics with the universality principles of physics addresses a longstanding gap and provides us with a deeper and more predictive understanding of population synchrony and ecosystem stability in complex environments.




\bibliography{main}

\end{document}